\documentclass[twocolumn,showpacs,preprintnumbers,amsmath,amssymb,prl]{revtex4}
\usepackage{amsfonts}
\usepackage[english]{babel}
\usepackage[T1]{fontenc}
\usepackage{times}
\usepackage{mathrsfs}
\usepackage{graphicx}% Include figure files
\usepackage{dcolumn}% Align table columns on decimal point
\usepackage{bm}% bold math
\usepackage{wasysym}
\usepackage[colorlinks,bookmarks=true,citecolor=blue,linkcolor=red,urlcolor=blue]{hyperref}
\usepackage[tight, FIGTOPCAP, hang, raggedright, nooneline]{subfigure}
\usepackage{hyperref}

\begin{document}
\title{Quantum oscillations without a Fermi surface and the anomalous de Haas-van Alphen effect}
\author {Johannes Knolle}
\author{Nigel R. Cooper}
\affiliation{T.C.M. Group, Cavendish Laboratory, J.~J.~Thomson Avenue, Cambridge CB3 0HE, United Kingdom}
\begin{abstract}
  The de Haas-van Alphen effect (dHvAe), describing oscillations of
  the magnetization as a function of magnetic field, is commonly
  assumed to be a definite sign for the presence of a Fermi surface
  (FS). Indeed, the effect forms the basis of a well-established experimental
  procedure for accurately measuring FS topology and geometry of
  metallic systems, with parameters commonly extracted by fitting to
  the Lifshitz-Kosevich (LK) theory based on Fermi liquid theory.
  Here we show that, in contrast to this canonical situation, there
  can be quantum oscillations even for band insulators of
  certain types. We provide simple analytic formulas describing the
  temperature dependence of the quantum oscillations in this setting, showing strong deviations from LK theory.  We draw connections to recent experiments
  and discuss how our results can be used in future experiments to
  accurately determine e.g. hybridization gaps in heavy fermion
  systems.
\end{abstract}

\date{\today}

\maketitle

{\it Introduction.} Landau quantization of electrons~\cite{Landau1930}, which leads to quantum oscillations (QO) of physical observables as a function of applied magnetic field~\cite{Alphen1930}, has been one of the cornerstones of condensed matter physics. On the one hand, it leads to new phenomena such as the integer quantum Hall effect~\cite{Klitzing1980} and its fractional version~\cite{Tsui1982}. For the latter, it even induces an unexpected new phase of matter beyond the standard Landau classification~\cite{Laughlin1983}, which ignited the field of topological phases~\cite{Wen2004}. On the other hand, it is itself an invaluable tool for the characterization of correlated metallic systems~\cite{Shoenberg1984}. The canonical LK~\cite{Lifshitz1956} theory of QO in metals showed that the periodicity, e.g. of the magnetization, is proportional to extremal cross sectional areas of the FS, thus turning QO into a precise quantitative and by now standard tool for determining FSs. In addition,  Lifshitz and Kosevich showed that it is possible to study correlation effects by extracting the effective mass, $m^*$, from the temperature dependence of the QO amplitudes given  by (for the first harmonic)
\begin{eqnarray}\label{LKnormal}
R_{\text{LK}}(T)= \frac{\chi}{\sinh \chi}  \  \  \text{with} \  \ \chi=\frac{2\pi^2 T}{\hbar \omega_c}
\end{eqnarray}
and the cyclotron frequency $\omega_c= \frac{eB}{m^*c}$.

Later the LK theory was extended to include more general self energy interaction effects~\cite{Luttinger1961,Engelsberg1970,Wassermann1989,Wasserman1996}, but these always preserved the general structure of the LK theory only renormalizing  parameters, e.g. $m^*$. 
It still comes as a great surprise that experimentally almost all materials, from weakly interacting metals to strongly correlated heavy fermion systems~\cite{Taillefer1987,Aoki2013,Li2014} or copper oxide high temperature superconductors~\cite{Taillefer2007,Sebastian2008,Vignolle2008,Sebastian2012,Barisic2013}, are consistent with a LK description which is manifestly an effective single particle theory.     There have been only very few exceptions for heavy fermion systems, e.g. CeCoIn$_5$~\cite{McCollam2005} and most recently the tentative topological Kondo insulator SmB$_6$~\cite{Sebastian2015}, violating the general temperature behaviour, Eq.~(\ref{LKnormal}). There have been recent theoretical studies on QO which explored novel effects due to symmetry breaking from commensurate~\cite{Carter2010,Chakravarty2012} or incommensurate~\cite{Zhang2015} charge density waves but they remained in the canonical LK framework. A notable exception is given by Ref.\onlinecite{Hartnoll2010} which derived a generalized formula for exotic quantum critical systems described via non-perturbative field theories.

%In this letter we establish that even effectively non-interacting band electrons can exhibit anomalous non-LK-like QO.

Historically, the firmly established understanding of QO is tied to the existence of a FS, which in principle impedes the following simple question: Can there be QO in an insulator? In this Letter we show that, surprisingly, the general answer is {\it yes}. This arises if the cyclotron frequency $\hbar \omega_c$ is of the order of the electronic gap and the band structure picks out a particular area of the Brillouin zone (BZ), as described below.   We further show that, even in this non-interacting setting, the electrons exhibit anomalous non-LK QOs.

We show that a simple band insulator of itinerant electrons hybridized with a localized flat band does  exhibit well-defined QO.
The periodicity is given by the area defined by the intersection of the unhybridized bands even if the chemical potential, $\mu$, is inside the hybridization gap or inside the flat part of the FS. In the latter case, the periodicity is equally unusual because it is not proportional to the FS area. We find that the temperature dependence of the oscillation amplitudes strongly differs from the standard LK theory: First,  if $\mu$ is inside the gap  QO amplitudes have a maximum at a temperature set by the hybridization gap, $\gamma$. Second, for a chemical potential inside the bands but close to the flat regions the behaviour is even more complex and governed by an additional energy scale, $\delta \mu$, which is the distance of $\mu$ above the bottom of the upper band. For $\delta \mu < 2\gamma \ll W$  there is a characteristic steep increase of the amplitudes towards lowest temperatures. 

Our main result is the general temperature dependence  
\begin{align}\label{LKnew}
R(T)& =  \chi \sum_{n=0}^{\infty} 2 e^{-2 \chi \left[n+\frac{1}{2} \right] \Gamma \left( \frac{\delta \mu}{\gamma},\frac{T}{\gamma},n\right)   }
\end{align}
which is calculated for a continuum model of our scenario with $\Gamma \left( \frac{\delta \mu}{\gamma},\frac{T}{\gamma},n\right) = 1+\left( \left[ \frac{2\delta \mu}{\gamma}\right]^2 +\left[\frac{4 \pi T}{\gamma} \left( n+\frac{1}{2} \right)  \right]^2 \right)^{-1}  $. A simple approximate formula 
\begin{equation}
R(t) \simeq R_0(T) = \frac{\chi}{\sinh \left( \chi \Gamma_0\right)} 
\end{equation}
is valid in the regime ${\hbar \omega_c} \gtrapprox 2\gamma$ or more generally for ${T} \gtrapprox 0.25\gamma$ where we can replace $\Gamma \to \Gamma_0 \equiv \Gamma \left( \frac{\delta \mu}{\gamma},\frac{T}{\gamma},n=0\right)$ to obtain a generalized LK-like form, which has a simple interpretation as a doping and also {\it temperature-dependent} effective mass renormalization.
In order to substantiate our unexpected findings we reproduce all our results in an unbiased numerical tight-binding lattice model calculation.

\begin{figure}[tb]
\begin{centering}
\includegraphics[width=1.0\columnwidth]{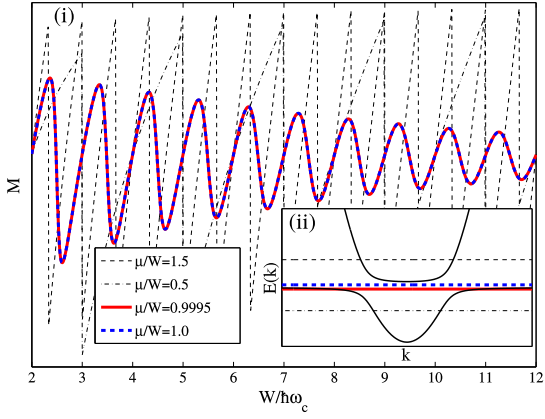}
\par\end{centering}
 \caption{(color online.) Main figure (i), Quantum oscillations of the magnetization, $M$, as a function of $W/\hbar \omega_c \propto 1/B$. Inset (ii), sketch of the band structure for our model (exaggerated hybridization gap for better visibility) and positions of the different chemical potentials $\mu$. If  $\mu$, is far away from the gap (black dashed and dot dashed), which is opened by hybridizing a localized flat band with an itinerant band, the periodicity of standard QO is proportional to the extremal cross section of the Fermi surface (here directly related to $\mu=\frac{S}{2\pi m}$ with the area $S=\pi k_F^2$). We find that even if $\mu$ is inside the gap (blue dashed) or in the flat band region (red) there are well defined QO which are directly proportional to the area picked out by the intersection of the unhybridized bands (here directly proportional to $W/\hbar \omega_c$).
\label{Fig1}}
\end{figure}

{\it The model.}
We consider non-interacting electrons with dispersion $\epsilon (\vec k)$ hybridized (strength $\frac{\gamma}{2}$) with a flat band of completely localized electrons at energy $W$. The microscopic origin of such a model is irrelevant for our discussion but the Kondo lattice model relevant for heavy fermion systems is effectively described by such a simple band structure at temperatures well below the Kondo temperature~\cite{Read1984,Auerbach1986,Millis1987}. The Hamiltonian is simply written as
\begin{align}
\label{Hamiltonian}
 H  =  \sum_{\vec k} 
\begin{bmatrix}
 \epsilon(\vec k) & \frac{\gamma}{2} \\
 \frac{\gamma}{2} & W
\end{bmatrix} 
\end{align}
with the two resulting energy bands 
$ E^0_{\pm}(\vec k)=\frac{1}{2} \left\lbrace \epsilon (\vec k) +W \pm \sqrt{\left(\epsilon (\vec k) -W\right)^2 +\gamma^2}\right\rbrace$
 separated by a hybridization gap $\gamma$ and centered around the flat band energy $W$ (blue dashed), see Fig.~(\ref{Fig1}) (ii). If  $\mu$ lies within the band gap  the system is insulating. 
Once an external magnetic field, $\vec B=B \vec z$, is switched on (described  by a vector potential $\vec A$) the Landau level (LL) structure is easily found for a continuum version of our model by replacing $\epsilon (\vec k)=\frac{1}{2 m} \left( \vec k -\frac{e}{c} \vec A\right)^2 \to      \hbar \omega_c\left(l+\frac{1}{2} \right)$ with $\omega_c =eB/m$, and
$\sum_{\vec k} \to N_\Phi \sum_l$ with $N_{\Phi}=\frac{BA}{\Phi_0}$ the number of  flux quanta $\Phi_0=\frac{hc}{2e}$ through the system area $A$. We have neglected the Zeeman energy splitting of spin components. For each LL index $l$ we have two energies with $E_-(l)<E_+(l)$ for all $l$. Note that for the lower band $E_-(l\to \infty) \to W$, giving a divergent density of states; 
this is an artefact of the continuum flat band which needs to be regularized. 

{\it Anomalous de Haas-van Alphen effect.}
We calculate the magnetization $M$ from the grand canonical potential ($k_B=1$)
\begin{eqnarray}\label{GCP}
M=-\frac{\partial \Omega}{\partial B}= \frac{\partial}{\partial B}  T \sum_{i} \ln \left[1 + e^{\frac{\mu-E_i}{ T}} \right] 
\end{eqnarray}
with a summation over all possible states including all degeneracies. 
We begin with the zero temperature behaviour 
\begin{eqnarray}
\label{PotentialFinal}
\Omega(\mu,T=0)= N_{\Phi}  \sum_{l,\pm; E_{\pm}(l)< \mu} \left\lbrace   E_{\pm}(l) -\mu\right\rbrace.
\end{eqnarray} 
We regularize the divergent sum over $E_-(l)$ by introducing a maximum chemical potential for that lower branch, $\frac{\mu_{\rm max}}{W} = \frac{1}{2} \left\lbrace n_{\rm max} +1 - \sqrt{\left[n_{\rm max}-1\right]^2+ \left[\frac{\gamma}{W}\right]^2 } \right\rbrace $ which is simply related to the maximum occupation $n_{\rm max}$ of the flat band without a field. Here, $n_{\rm max}$ is defined relative to the filling of a dispersive band $\epsilon(\vec k)$ with Fermi energy $\mu=W$ which defines an occupied area of the BZ $S$. For our continuum model with $\epsilon(\vec k) =\frac{k^2}{2m}$ we simply have $S=\pi k_F^2$ and  the relation $\mu=\frac{S}{2 \pi m}$ straightforwardly generalizes our results to general dispersions $\epsilon(\vec k)$~\cite{Shoenberg1984}.

In Fig.~\ref{Fig1} (i) we show the variation of $M$ as a function of magnetic field for different chemical potentials (fixed $\gamma/W=0.05$, $n_{\rm max}=5$ and all our findings are independent of the cut-off occupation $n_{\rm max}$). For $\mu$ far above (below) the gap there are the usual sharp QO with periodicity $\frac{\mu}{\hbar \omega_c}$ directly proportional to the occupied FS volume, see the black dashed (dot dashed) curves. For $\mu$ inside the gap (blue dashed) or inside the flat part of the bands (red) we still find well defined anomalous QO of comparable amplitudes. However, now these QO have a periodicity   $\frac{W}{\hbar \omega_c}$, hence a BZ area defined by the intersection of the unhybridized bands! For larger values of $\gamma/W$ (not shown) the amplitude of QO are strongly suppressed for smaller magnetic fields but as long as $\hbar \omega_c \gtrsim \gamma$ they remain observable.

{\it Effect of temperature.}
Next, we study the temperature dependence which can by easily calculated for free electrons from $\Omega(\mu,T=0)$ via the convolution~\cite{Shoenberg1984}
\begin{eqnarray}
\Omega(\mu,T)= - \int_{-\infty}^{\infty} \frac{\partial f(\xi-\mu)}{\partial \xi} \Omega(\xi,0) \text{d}\xi 
\label{FiniteT}
\end{eqnarray}
with the derivative of the Fermi function $-\frac{\partial f(\xi-\mu)}{\partial \xi}=\frac{1}{2 T \left[ 1+\cosh \frac{\xi-\mu}{ T}\right] }$ which is strongly peaked at $\xi=\mu$ with a width set by temperature. The advantage of this expression is its intuitive interpretation: it is a weighted average over different chemical potentials from a window proportional to temperature. For standard QO different $\mu$ correspond to different periods, hence increasing $T$ always damps the sharp amplitudes via dephasing. Evaluating Eq.~(\ref{PotentialFinal},\ref{FiniteT}) numerically, we find that this is not the case for our system: e.g. for $\mu=W$ inside the gap we find an initial {\it increase} of the amplitudes up to a maximum at $T\approx \gamma/4$ before damping sets in (not shown). This arises because in the temperature average over different $\mu$ all QO have the same periodicity (at least for low $T$) preventing dephasing, however those from regions in the flat part have a larger amplitude.  

\begin{figure}[tb]
\begin{centering}
\includegraphics[width=1.0\columnwidth]{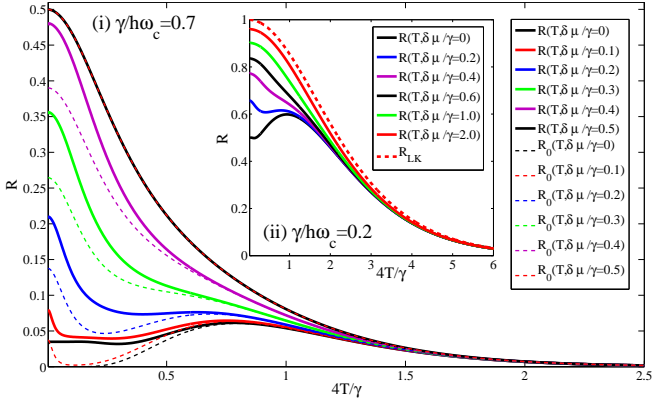}
\par\end{centering}
 \caption{(color online.) Temperature dependence of the damping factor R(T). In (i) it is shown for $\frac{\gamma}{\hbar \omega_c}=0.7$ and for different values of the chemical potential, $\mu=W=\delta \mu$, parametrized by $\frac{\delta \mu}{\gamma}$. Dashed lines are calculated from the approximate $R_0(T)$ in which $\Gamma$ is replaced by $\Gamma_0=\Gamma(n=0)$. The inset (ii) shows the same for  a different value $\frac{\gamma}{\hbar \omega_c}=0.2$. In this case $R_0(T)$ always coincides with the exact $R(T)$. Note that for large values of $\frac{\delta \mu}{\gamma}$ the standard Lifshitz-Kosevich behaviour $R_{\rm LK}$ (red dashed) is quickly approached. 
\label{Fig2}}
\end{figure}
For an analytical calculation of the $T$-dependence we follow earlier work~\cite{Hartnoll2010,Wasserman1996} using a finite temperature description in terms of Matsubara frequencies $\omega_n=2 \pi i T (n+\frac{1}{2})$. The oscillatory part of the grand canonical potential takes the form
$\Omega(\mu,T)= \hbar N_{\Phi} \sum_{k=1}^{\infty} \frac{1}{k} \mathfrak{Re} \sum_{n=0}^{\infty}  e^{i 2 \pi k l^* (n)}  $
with $l^*(n)$ being the LL index which defines the pole of the Greens function $G(i\omega_n,l)=\left(i \omega_n -\left[ E_{\pm}(l)-\mu \right]  \right)^{-1} $. We write $\mu=W+\delta \mu$ and find a single $l^*$ to obtain 
\begin{align}
\Omega(\mu,T)= \hbar N_{\Phi} \sum_{k=1}^{\infty} \frac{1}{k} \cos \frac{2 \pi k W}{\hbar \omega_c}  \sum_{n=0}^{\infty}  e^{- \frac{4 \pi^2 k T (n+\frac{1}{2})}{\hbar \omega_c} \Gamma\left(\frac{\delta \mu}{\gamma}, \frac{T}{\gamma},n\right)  }  
\label{Omeg_osc1}
\end{align}
where we have neglected a small $n$- and $\delta \mu$-dependence of the real part of $l^*$ which only slightly modifies the periodicity but not the damping; $\Gamma\left(\frac{\delta \mu}{\gamma}, \frac{T}{\gamma},n\right)$ is defined below Eq.~(\ref{LKnew}). Now differentiating w.r.t. magnetic field and in the limit $\frac{\delta \mu}{W},\frac{\hbar \omega_c}{W}\ll 1$ we obtain the final result for the  first harmonic $k=1$ of the magnetization:
\begin{align}
M=  -\frac{A W e}{2 \pi^2 c } \sin \frac{2 \pi W}{\hbar \omega_c} R(T) 
\label{M_osc}
\end{align}
with the damping factor $R(T)$ given in  Eq.~(\ref{LKnew}).

In Fig.~\ref{Fig2} we plot representative curves of $R(T)$ which fully capture the behaviour we have found by numerically evaluating Eq.~(\ref{PotentialFinal},\ref{FiniteT}). For a chemical potential inside the gap ($\frac{\delta \mu}{\gamma}=0$ black curves) there is an increase of the amplitudes up to a maximum $T$ which is set by the energy scale of the hybridization $\gamma$ itself. The total (relative) height of the maximum increases (decreases) for smaller $\frac{\gamma}{\hbar \omega_c}$ [see inset (ii)]. For larger or smaller fillings a characteristic steep increase of the amplitude at a scale $T\approx \delta\mu/10$ is observed. The simple approximate formula $R_0(T)$, see Eq.~(\ref{LKnew}), in general reproduces the behaviour of $R(T)$ for sufficiently large temperatures [dashed curves in (i)]. For small values of $\frac{\gamma}{\hbar \omega_c}$ it fully captures the exact result as shown in the inset (ii).

\begin{figure}[tb]
\begin{centering}
\includegraphics[width=1.0\columnwidth]{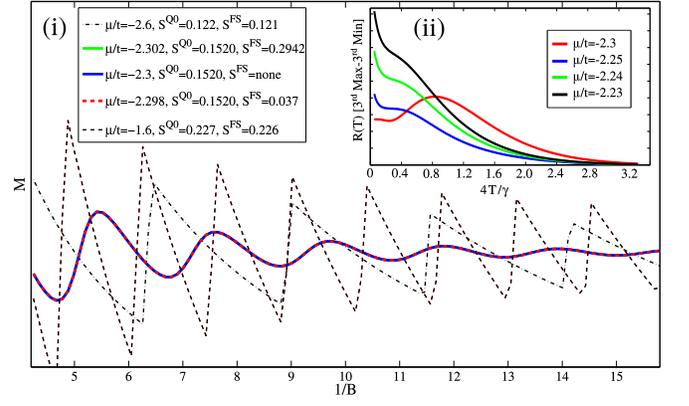}
\par\end{centering}
 \caption{(color online.) QO oscillations for a tight-binding lattice model with $W=-2.3$ and $\gamma=0.2$ (all energies in units of $t$). In the main figure (i) for $T=0$ it is confirmed that as long as the chemical potential is far away from the gap (black dashed and dot dashed) the FS area, $S^{\text{QO}}$ extracted from the QO period nicely reproduce the area $S^{\text{FS}}$ obtained from the relative BZ area of the FS for zero field, see legend. If $\mu$ is inside or close to the gap (green, blue, red dashed) we find anomalous QO as before with a periodicity not related to $S^{\text{FS}}$. In the inset (ii), we extracted the temperature dependence $R(T)$ by calculating the difference between a consecutive minimum and maximum of $M$ as a function of temperature, which confirms the analytical behaviour of Eq.~(\ref{LKnew}), compare to Fig.~\ref{Fig2}.   
\label{Fig3}}
\end{figure}
{\it Lattice model.}
So far our theory was restricted to a continuum description, requiring regularization of the flat band occupation. 
To confirm our findings for a microscopic model, we have performed a full lattice tight binding calculation. We consider a model of spinless electrons  on a square lattice, with Hamiltonian
\begin{eqnarray}
H 
& = & - \sum_{\langle i,j\rangle} 
\left( 
t_{ij} \hat{c}^\dag_i\hat{c}_j +  \mbox{h.c.} 
\right)
\nonumber \\
 & &  
%\left[
+ \frac{\gamma}{2} 
\sum_i 
\left(\hat{c}^\dag_i\hat{f}_i + \mbox{h.c.} \right)
+ \sum_i W \hat{f}^\dag_i \hat{f}_i
%\right]
\end{eqnarray}
The magnetic field
is incorporated in the phases of the nearest-neighbour hopping parameters $t_{ij}$ via the usual Peierls substitution.
These itinerant electrons are coupled locally to a second completely localized orbital with on-site energy $W$ at each site. 
The magnetic flux through the magnetic unit cell of size $L_x L_y$ is quantized to multiples of the elementary flux quantum  $\Phi =L_x L_y B = m \Phi_0$. We study the system at a series of magnetic fields for which $L_y=2$ and there is an integer $L_x$ such that the flux $\Phi=\Phi_0$.
For each  field the Hamiltonian is easily diagonalized as before, but now the maximum occupation of the flat band is fixed by the total number of lattice sites.
The QO are directly calculated from the grand canonical potential, Eq.~(\ref{GCP}).
%Note, it turns out from the numerics that for $t_x=t_y$ the QO are basically independent of the number of discretization steps in the $k_y$ direction.

In Fig.~\ref{Fig3} we show the QO of the magnetization which we obtain from our lattice simulation. We not only recover the anomalous dHvAe at $T=0$, see main panel (i), but we also confirm the peculiar temperature dependence $R(T)$ of the amplitudes, see inset (ii). If the chemical potential lies in the flat part of the band such that $\delta \mu\ll W$ we recover the peculiar upturn of the amplitudes towards the lowest $T$.

{\it Discussion and conclusion.}
We have shown that, at odds with the canonical understanding of QO in metals, a simple model of itinerant electrons coupled to a flat band can lead to clear QO even in the complete absence of a FS. We find strong deviations of the  temperature dependence from the usual LK theory and derived analytic expressions which can be tested in future experiments. 
We believe that our results are most promisingly applicable to certain heavy fermion materials whose properties well below the Kondo temperature are effectively described by a band structure similar to our model~\cite{Read1984,Auerbach1986,Millis1987}. In that context it is worth pointing out that our theory has its most prominent deviations from the LK description in a regime in which the cyclotron frequency, $\hbar \omega_c$, is larger than the hybridization strength $\gamma$ as well as the activation gap $\gamma^2/4W$ 
-- a condition fulfilled at least by some heavy fermion materials.  

Interestingly, the main features of our peculiar temperature dependence were already observed in heavy fermion compounds in two of the rarely available experimental examples deviating from LK theory: Amplitudes of some frequencies of the dHvAe in CeCoIn$_5$ display a clear maximum at a nonzero temperature of 100 mK, which has been attributed to a fine tuned spin-dependent mass enhancement. Most recently, the tentative topological Kondo insulator SmB$_6$~\cite{Dzero2010}, for which the appearance of QO itself despite the opening of an activation gap~\cite{Li2014}  (as seen in transport) has been a puzzle, does show QO with a very strong increase of intensity below 1K signaling the presence of a second low energy scale in the system~\cite{Sebastian2015}. Although, the latter is likely an interaction effect it is interesting to note that in our non-interacting theory a chemical potential not in the gap but just touching one of the heavy bands ($|\delta \mu/\gamma|>0$)~\cite{Golden2013} sets a new energy scale and gives a very similar temperature dependence with a steep increase of the amplitudes at very low temperatures, see Fig.~\ref{Fig2}. For the actual material SmB$_6$ it is more likely that our scenario just explains why there are QO in this Kondo insulating system at all but the incorporation of self energy effects into our theory, which will introduce a second energy scale from coupling to collective modes, is a promising route for future investigations.     
In addition, it is an open question for future research, whether certain semiconductors with small direct band gaps could also display similar anomalous dHvAes.

Despite many decades of intense research on the dHvAe we have demonstrated that it still holds surprises -- there can be be QO even in insulating systems. Beyond a mere curiosity the interest in standard LK-like QO derives from its capacity of accurately determining FSs. Similarly, we anticipate  that our {\it anomalous dHvAe}  applicable to heavy Fermi liquids will be useful in the future for determining hybridization gaps (proportional to the Kondo coupling) by measuring the temperature of maximum amplitudes.

{\it Acknowledgments.} 
We thank D. Khmelnitskii for discussion.
It is a pleasure to acknowledge helpful discussions with G. Lonzarich and S. Sebastian and for sharing their experimental data on SmB$_6$ prior to publication~\cite{Sebastian2015}. The work of J.K. is supported by a Fellowship within the Postdoc-Program of the German Academic Exchange Service (DAAD).

\vspace{2.5cm}
\section{Supplementary materials}
Here, we present details of our tight-binding lattice model calculation. We use a model of spinless electrons in a uniform magnetic field on the square lattice. In addition to itinerant delocalized electrons with a dispersion $\epsilon (\vec k)=-2t \cos (k_x)-2 t \cos (k_y)$ there is a second completely localized orbital with on-site energy $W$ at each site. Both d.o.f. are  hybridized and still described by Eq.~(\ref{Hamiltonian}). The system is diagonalized as before but now the maximum occupation of the flat band is fixed by the total number of lattice sites.
A magnetic field corresponding to the discrete vector potential $\vec A = B x \vec y$  is incorporated into the tight-binding hopping parameter $t$ via the usual Peierl's substitution $t_{x,y \to x',y'} \to e^{i \int_{x',y'}^{x,y} \vec A \cdot \text{d} \vec r} t_{x,y \to x',y'}$.
The magnetic flux through the magnetic unit cell of size $L_x L_y$ is quantized to multiples of the elementary flux quantum  $\Phi =L_x L_y B = m \Phi_0$. We work in a gauge with fixed $L_y=2$ and $m$ such that we have a one dimensional unit cell of length $L$ proportional to $B$. (We put $L_x=L$ and use an elementary flux quantum with units $\Phi_0=2\pi$.)

In the following, the upper index $l$ labels the position inside each unit cell and $( x,  y)$ label the position of the unit cells. Itinerant (localized) electrons  are created by operators $\hat c^{l\dagger}_{x,y}$ ($\hat f^{l\dagger}_{x,y}$). The lattice Hamiltonian takes the form
$H =  -\sum_{x,y}  t   \left[ \sum_{l=0}^{L-2} \hat c^{l+1\dagger}_{x,y} \hat c^l_{x,y} + \hat c_{x+L,y}^{0\dagger}\hat  c_{x,y}^{L-1} +h.c. \right] - \\ \nonumber
    \sum_{x,y} \sum_{l=0}^{L-1} \left[t  e^{i l \Phi }\hat c^{l \dagger}_{x,y+1} \hat c^l_{x,y}+ \frac{\gamma}{2} \hat f^{l\dagger}_{x,y} \hat c^l_{x,y}  + h.c. \right] - \\  \sum_{x,y} \sum_{l=0}^{L-1}  \left [W \hat f^{l\dagger}_{x,y} \hat f^l_{x,y} \right]$.
For each field $B$ we have a translationally invariant system with unit cells of length $L$. Then, we use a Fourier transform $\hat c_{x,y}^{l}=\frac{1}{\sqrt{N}} \sum_{k_x,k_y} e^{i k_x x -i k_y y }\hat  c_{k_x,k_y}^l$ and a corresponding spinor
$\Psi_{\mathbf{k}}^\dagger = [\hat c_{\mathbf{k}}^{0 \dagger}, \hat c_{\mathbf{k}}^{1 \dagger} , \dots , \hat c_{\mathbf{k}}^{L-1 \dagger}, \hat f_{\mathbf{k}}^{0 \dagger}, \hat f_{\mathbf{k}}^{1 \dagger}, \dots, \hat f_{\mathbf{k}}^{L-1 \dagger}]$ such that the energies are easily found from the resulting $2L \times 2L$ quadratic form.
\begin{eqnarray}
\label{H_bigmatrix}
H& =\sum_{\mathbf{k}} \Psi_{\mathbf{k}}^{\dagger} 
\begin{pmatrix}
 \hat H_c & \hat H_{\gamma} \\
 \hat H_{\gamma} & \hat H_W
\end{pmatrix}
\Psi_{\mathbf{k}}
\end{eqnarray}
with the matrix
\begin{widetext}
\begin{eqnarray}
\hat H_c =
\begin{pmatrix}
- 2 t_y \cos( k_y - 0 \cdot \Phi ) & -t_x & \cdots & e^{-ik_x L} \\
-t_x & - 2 t_y \cos( k_y - 1 \cdot \Phi) & - t_x &  \vdots \\
\vdots & -t_x & \ddots & -t_x \\
e^{ik_x L} & \cdots & -t_x & - 2 t_y \cos( k_y - (L-1)\cdot \Phi) 
\end{pmatrix} 
\end{eqnarray}
\end{widetext}

and the diagonal $L\times L$ matrices $\hat H_{\gamma}= \frac{\gamma}{2}  \hat{\mathbf{1}}$ and $\hat H_{W}= W  \hat{\mathbf{1}}$.
Finally, QO are directly calculated from the grand canonical potential, Eq.~(\ref{GCP}).

\end{document}